***KILLER TECHNOLOGIES:***
**THE DESTRUCTIVE CREATION IN THE TECHNICAL CHANGE**

*Mario Coccia[1]*
CNR -- National Research Council of Italy & yale university

CNR -- National Research Council of Italy, Collegio Carlo Alberto, Via Real Collegio, 30-10024 Moncalieri (Torino, Italy)

Yale school of medicine, 310 Cedar Street, Lauder Hall, New Haven, CT 06510, USA

*E*-mail: mario.coccia@cnr.it

**Abstract**. *Killer technology* is a radical innovation, based on new products and/or processes, that with high technical and/or economic performance destroys the usage value of established techniques previously sold and used. Killer technology is a *new concept* in economics of innovation that may be useful for bringing a new perspective to explain and generalize the behavior and characteristics of innovations that generate a destructive creation for sustaining technical change. To explore the behavior of killer technologies, a simple model is proposed to analyze and predict how killer technologies destroy and substitute established technologies. Empirical evidence of this theoretical framework is based on historical data on the evolution of some example technologies. Theoretical framework and empirical evidence hint at general properties of the behavior of killer technologies to explain corporate, industrial, economic and social change and to support best practices for technology management of firms and innovation policy of nations. Overall, then, the proposed theoretical framework can lay a foundation for the development of more sophisticated concepts to explain the behavior of vital technologies that generate technological and industrial change in society.

**Keywords**: Killer Technology; Radical Innovation; Destructive Creation; Evolution of Technology; Technological Evolution; Nature of Technology; Technology Change; Innovation Management.

**JEL codes:** O30; O32; O33; B50

---

[1]      Acknowledgement. I gratefully acknowledge financial support from National Research Council of Italy–Direzione Generale Relazioni Internazionali for funding this research project developed at Yale University in 2019 (grant-cnr n. 62489-2018). The author declares that he has no relevant or material financial interests that relate to the research discussed in this paper.

**Introduction and goals of the investigation**

This paper has three goals. The first is to define the concept of killer technology, a new perspective that may explain and generalize vital elements of technological change in turbulent markets. The second goal is to propose a model and provide an empirical evidence based on historical data of example technologies to analyze the behavior and characteristics of killer technologies. Finally, the third goal is to suggest general properties that can explain and generalize the behavior of killer technologies for sustaining industrial and economic change in society.

This study is part of a large body of research on the evolution of technology to explain, with a new perspective, technological, economic and social change (Coccia, 2017, 2018, 2019). In the research field of technological evolution, Hosler (1994, p. 3, original italics) argues that the development of technology is, at least to some extent, influenced by "technical *choices*", which express social and political factors, and "technical *requirements*", imposed by material properties. In this context, Arthur and Polak (2006, p. 23) claim that: "Technology … evolves by constructing new devices and methods from ones that previously exist, and in turn offering these as possible components—building blocks—for the construction of further new devices and elements". Calvano (2006) explains the role of specific technologies in technical change with the concept of "destructive creation", in which "a monopolist has the option, at the beginning of each period, to destroy the usage value of all units previously sold and simultaneously introduce a new, perhaps improved, vintage at some cost $c \geq 0$…Such cost is interpreted as any expenditure incurred in the process of destruction as well as in the process of creating, developing and marketing the new versions". In fact, technical change, according to Pistorius and Utterback (1997), can be also due to a rivalry between technologies in a context of competitive markets in which emerging technologies often substitute for more mature technologies.

Although several contributions in these fields of research, the behavior and characteristics of specific typologies of technological innovations that generate the radical change in markets and technical change in society are hardly known.

This study proposes a new concept in economics of innovation, the *killer technologies* that generate a disruptive creation in a Schumpeterian world oriented to continuous technological, economic and social change. Hence, the main aim of this article is to explain and generalize whenever possible, the behavior and characteristics of killer technologies within industrial competition. In particular, this study addresses some basic questions: what are the *degree* and *rate* at which new killer technologies are adopted when they attempt in substituting for existing victim technologies? What are the *properties* of killer technologies in a setting of competition between technologies in markets? And finally, what are the *consequences* of killer technologies for technical change?

Next sections endeavor to explain how a specific typology of radical innovation, called *killer technology*, affects other technologies and generate corporate, industrial and economic change in society.

**Theoretical framework**

Arthur (2009, p. 15ff) claimed that one of the most important problems to understand regarding technology is to explain how it evolves and generates technical change (cf., Arthur and Polak, 2006; Basalla, 1988). Technological evolution can be explained in economics of technology with theories based on processes of competitive substitution of a new technology for the old one (Fisher and Pry, 1971; Sahal, 1981). Theories of competitive substitution between technologies state that the adoption of a new technology is associated with the nature of some comparable older technology in use, such that an established technology improves when confronted with the prospect of being substituted by a new technology (Sahal, 1981; Utterback et al., 2019). In particular, when comparable technologies do exist, each technology tends to affect the behavior and evolutionary pathway of other technologies (Coccia, 2018). Pistorius and Utterback (1997) argue that new technologies often supplant for more mature technologies in markets. This interaction between technologies is usually referred to as competition that leads to the dominance of a technology on another one in turbulent markets (cf., Berg et al., 2019; Moehrle and Caferoglu, 2019).

A model that operationalizes the competition between technologies was suggested by Fisher and Pry (1971). This model proposes that the evolution of a new product/process as a substitute for a prior one can be plotted in the form of $f / (1 - f)$ as a function of time on a semilogarithmic graph, generating a straight line through the resulting points ($f$ = market share of the emerging product versus time; cf., Fisher and Pry, 1971, p. 77). Moreover, if data on the absolute adoption of a new technique relative to the use of the old technique are plotted on double-logarithmic paper, the resulting trend is also approximately linear (Sahal, 1981). Fisher and Pry (1971) show that substitution models fit to data of competition between technologies, such as synthetic vs. natural fibers, synthetic vs. natural rubber, etc. In general, technological advances are given by competitive substitutions of one artifact satisfying a need for another. Fisher and Pry (1971, p. 88) state that: "The speed with which a substitution takes place is not a simple measure of the pace of technical advance . . . . it is, rather a measure of the unbalance in these factors between the competitive elements of the substitution".

The competition between technologies can also generate a predator-prey relation, where one technology enhances the growth rate of the other but the second inhibits the growth rate of the first (Pistorius and Utterback, 1997, p. 74). In particular, a predator-prey relationship can exist in the presence of competition between an emerging technology and a mature technology in a niche market. In this case, emerging technology will benefit from the presence of mature technology. At the same time, emerging technology may slowly reduce market share of mature technology. In this context, Pistorius and Utterback (1997, p. 72) argue that: "Pure competition, where an emerging technology has a negative influence on the growth of a mature technology, and the mature technology has a negative influence on the growth of the emerging technology". Farrell (1993) used a model based on Lotka-Volterra equations to examine this competition between technologies, such as nylon versus rayon tire cords, telephone versus telegraph usage, etc. Utterback et al. (2019) show a predator-prey relation in a specific period between plywood and Oriented Strand Board (OSB is a composite of oriented and layered strands, peeled from widely available smaller trees). In short, on the one hand, a predator-prey interaction has emerging technology in the role of predator and the mature technology as the prey. On the other hand, one can also visualize a situation where the mature technology is the predator and the emerging technology is the prey (Pistorius and Utterback, 1997, p. 78).

In general, competition is often embodied in substitutes, which have a powerful force in markets to improve products and processes and generate technical change. Porter (1980) considers substitutes as one of the five forces of industrial competition. These approaches oriented to competition between technologies seem to be appropriate to explain technological advances of specific product and process innovations in turbulent markets. In this research field, the study here proposes a new concept, the killer technology, that seeks to explain the behavior and characteristics of specific radical innovations in the dynamics of industrial competition. In particular, the behavior of killer technologies is especially relevant to explain how a new technology destroys established technologies, enhances dynamic capabilities and competitive advantage of leading firms and generates technical change in society (cf., Teece et al., 1997; Porter, 1980).

**Definition, examples and evolutionary model of killer technologies**

The primary goal of this study is to define the concept of killer technology; and that definition should meet the conditions of independence, generality, epistemological applicability and empirical correctness (Brandon, 1978). The following premises support the proposed theory here:

a) Technology is a complex system of artifact that is composed of more than one element

and/or sub-system and a relationship that holds between each element and at least one other element in the system. Technology, produced and used by living systems, is selected and adapted in Environment $E$ (such as market), considering technical and economic characteristics to satisfy needs, achieve goals and/or solve problems in society.

b) Radical innovations are the result of a research and development activity (in firms, universities and/or government labs) that generates a discontinuous change in the evolutionary pathway of technologies, affecting the growth of a sector or giving rise to new sectors. Radical innovations of product are for instance contraceptive pills, smartphones, contact lens, etc., whereas radical innovations of process are oxygen steelmaking process, Solvay process, etc. Radical innovations generate big improvements in the cost and quality of products and/or processes to satisfy needs of users and/or solve problems in society.

c) In the long run, the behavior and evolution of any technology is *not independent* from the behavior and evolution of other technologies (Coccia, 2018, 2018a).

---

*Definition of killer technology*

Killer technology is a radical innovation, based on new products and/or processes, that with high technical and/or economic performance destroys the usage value of all established techniques or technological devices previously sold, generating improvements in technical choices, costs and quality to sustain competitive advantage of firms, satisfy needs of people and/or solve problems in society.

---

*Remark:* a killer technology in the maturity phase of the cycle of development can change its status in victim technology because of new technologies that, in turn, become killer technologies.

o     *Examples of killer technologies in the history of technology*

Sahal (1981, p. 79ff) explains the diffusion of steamship and sailing ship from 1850s. The competition between steamship and sailing ship generates in the first phase an improvement of sailing ships by a number of incremental innovations (Graham, 1956). However, steamship in the long run has sequential radical technological advances based on substitution of the screw propeller for the paddle wheel, the development of compound engine, the application of steel in place of iron, the adoption of high pressure triple expansion engine that reduces the fuel consumption of steamships and increases the speed of service, etc. (Gilfillan, 1935). This competition generates in the long run a dominance of steamships, as killer technologies, over sailing ships as means of transportation of goods and people (cf., Rosenberg, 1976).

Another main example of killer technology is the diffusion of Solvay process that in the 1900s destroys the Leblanc process in the production of soda. In particular, the competition between these innovations generates, in the long run, vital technological advances of Solvay process and the advent of this new process technology in the manufacturing sector of soda (Freeman, 1974).

In agriculture, the plowing is one of the most energy-consuming operations (Walker, 1929). The farm tractor is a killer technology that generates a substitution of mechanical for animal power. In fact, farm tractor is a general-purpose technology in agriculture that can be applied for plowing and a wider range of farm operations (Sahal, 1981).

A final example is storage devices. Sony corporation introduced in 1983 micro diskettes: 3.5-inch floppy disks that remained a popular medium of storage for many years, but they decline by the mid-1990s (Coccia, 2018b; Mee and Daniel, 1996). The development of a new storage device based on Universal Serial Bus (USB) technology began in 1995 by Intel to standardize the connection of computer peripherals (Coccia, 2018b). The USB 1.0 in 1995 transferred data at a rate of 12 megabits (MB/s) per second. This new technology in interaction with host technologies, such as Personal Computers (PCs), destroys the markets of floppy disks because of more efficient operations of storage, higher velocity of transfer data (in USB 3.0 is about 800 MB/s) and of storage capacity up to 4TB in 2019 for portable storage (Coccia, 2018, 2018b, 2019). In 1998, the Personal Computer iMac G3 by Apple Inc. was the first consumer computer to discontinue legacy ports (serial and parallel) in favor of USB technology (Coccia, 2018b). This innovation strategy by Apple Inc., a market leader, helped to pave the way for a market of solely USB peripherals rather than other ports for storage devices, such that USB devices and other portable storage, in the role of killer technologies, have destroyed the use of 3.5-inch floppy disks, Compact Disc, etc., generating a market shift and industrial change (Coccia, 2018).

o    *Proposed evolutionary model of killer technology*
The second goal of this study is to operationalize the behavior of killer technology *vs.* victim technology proponing a simple model of technological growth of a killer technology *Kl* (a new radical technology) in relation to a victim technology *V* (established technology). This approach is based on the biological principle of allometry that was originated in zoology to study the differential growth rates of the parts of a living organism's body in relation to the whole body (cf., Reeve and Huxley, 1945). Sahal (1981) applies this model to explain patterns of technological innovation with interesting results for spatial diffusion of technology.

The general model here is based on following assumptions.

(1)  Suppose the simplest possible case of only two technologies, $V$ (*victim technology* or established technology) and $Kl$ (a *killer technology* or new technology).

(2)  Let $Kl(t)$ be the level of a killer technology $Kl$ at the time $t$ and $V(t)$ be the level of a victim technology $V$ at the same time.

Suppose that both $Kl$ and $V$ evolve according to some $S$-shaped pattern of technological growth, such a pattern can be represented analytically in terms of the differential equation of logistic function. For $V$, victim technology, the starting equation is:

$$\frac{1}{V}\frac{dV}{dt} = \frac{b_1}{K_1}\left(K_1 - V\right)$$

The equation can be rewritten as:

$$\frac{K_1}{V}\frac{1}{\left(K_1 - V\right)}dV = b_1 dt$$

The integral of this equation is:

$$\log V - \log\left(K_1 - V\right) = A \; + b_1 t$$

$$\log\frac{K_1 - V}{V} = a_1 - b_1 t$$

$$V = \frac{K_1}{1 + \exp\left(a_1 - b_1 t\right)}$$

$a_1 = b_1 t$ and $t$ = abscissa of the point of inflection.

The growth of $V(t)$ can be described respectively as:

$$\log \frac{K_1 - V}{V} = a_1 - b_1 t \qquad\qquad [1]$$

*Mutatis mutandis*, for killer technology $Kl(t)$ the equation is:

$$\log \frac{K_2 - Kl}{Kl} = a_2 - b_2 t \qquad\qquad [2]$$

The logistic curve here is a symmetrical *S*-shaped curve with a point of inflection at 0.5K with $a_{1,2}$ are constants depending on initial conditions, $K_{1,2}$ are equilibrium levels of growth, and $b_{1,2}$ are rate-of-growth parameters (1= victim technology: $V$; 2= killer technology: $Kl$).

Solving equations [1] and [2] for $t$, the result is:

$$t = \frac{a_1}{b_1} - \frac{1}{b_1} \log \frac{K_1 - V}{V} = \frac{a_2}{b_2} - \frac{1}{b_2} \log \frac{K_2 - Kl}{Kl}$$

The expression generated is:

$$\frac{V}{K_1 - V} = C_1 \left( \frac{Kl}{K_2 - Kl} \right)^{\frac{b_1}{b_2}} \qquad\qquad [3]$$

Equation [3] in a simplified form is $C_1 = exp[b_1(t_2 - t_1)]$ with $a_1 = b_1 t_1$ and $a_2 = b_2 t_2$ (cf. Eqs. [1] and [2]); when $Kl$ and $V$ are small in comparison with their final value, the model of evolutionary growth of killer technology in relation to victim technology is given by:

$$Kl = A \ (V)^B \qquad\qquad [4]$$

where $A = \dfrac{K_2}{(K_1)^{\frac{b_2}{b_1}}} C_1$ and $B = \dfrac{b_2}{b_1}$

The logarithmic form of the equation [4] is a simple linear relationship:

$$\log Kl = \log A + B \ \log V \qquad\qquad [5]$$

$B$ is the coefficient of growth that measures the evolution of technology $Kl$ (killer) in relation to $V$ (victim technology).

This model of the evolution of killer technology [5] has linear parameters that are estimated with the Ordinary Least-Squares Method. The value of $B \gtrless 1$ in the model [5] measures the relative growth of $Kl$ in relation to the growth of $V$ and it indicates different patterns of technological evolution:

In particular,

- $B < 1$, whether technology $Kl$ destroys at a lower relative rate of change victim technology over the course of time (*under-development of killer technology*).

- $B$ has a unit value: $B = 1$, then the killer technology $Kl$ kills and substitutes victim technology at a proportional rate of change (*proportional growth of killer technology*).

- $B > 1$, whether killer technology $Kl$ kills victim technology at greater relative rate of change over the course of time (*development of killer technology*).

Overall, then, the coefficient $B$ of growth can be a metric for analyzing the behavior of growth of killer technology in relation to victim technology in markets.

## Materials and methods

▪ *Data and their sources*

The analysis of killer technology is measured here using historical data of four example technologies (three for US market and one for Canada market):

o   Farm tractor in the USA, 1920-1960 period

o   Hydro-and Thermoelectric power in Canada, 1917-1972

o   Diesel-powered tractors in the USA, 1955-1971

o   Technologies for recorded music in the USA, 1973-2018 (Cassette, CD and streaming technology)

US and Canadian national systems of innovation are vital cases study to show general patterns of the evolution of technology across advanced economies (Steil et al., 2002). Sources of data for three technologies are tables published by Sahal (1981, pp.319-350, originally sourced from trade literature). In the case of recorded music technology, the source is Recording Industry Association of America (RIAA) which provides data on U.S. recorded music revenues and shipments dating all the way back to 1973 (RIAA, 2019). Note that data from the earliest years and also the war years are sparse for some technologies. Moreover, in all of these examples, the first year represented is not the year of invention (cf., Sahal, 1981; RIAA, 2019).

▪ *Measures*

1. Farm tractor in the USA, 1920-1960 period

   − Growth in the number of tractors on farms in thousands (mechanical power): killer technology (*Kl*)

   − Number of horses on farms in thousands (animal power): victim technology (*V*)

2. Hydro-and thermoelectric generating units in Canada, 1917-1972. In particular, thermal power is studied in relation to the growth of hydroelectric power. Note that the growth of thermal power reflects the diffusion of both fossil fuel and nuclear power units (Sahal, 1981, p. 91). The specific measures of this technology are given by:

   − Thermoelectric power in installed capacity in megawatts (MW): killer technology (*Kl*)

   − Hydropower in installed capacity in megawatts (MW): victim technology (*V*)

3. Diesel-powered tractors in the USA, 1955-1971 period

   − Annual production of diesel-powered tractors: killer technology (*Kl*)

   − Annual production of gasoline powered tractors: victim technology (*V*)

4. Cassette, CD and streaming technologies for recorded music in the USA, 1973-2018 period

*First phase*:

− Recorded music revenues in millions $ (adjusted for inflation, 2018 Dollars) of CD as killer technology (*Kl*)

− Recorded music revenues in millions $ (adjusted for inflation, 2018 Dollars) of Cassettes as victim technology (*V*)

*Second phase:*

− Recorded music revenues in millions \$ (adjusted for inflation, 2018 Dollars) of streaming as killer technology (*Kl* ). Note that streaming technology is measured here including different modes: paid subscription, on-demand streaming, other Ad-supported streaming, sound exchange distributions and limited tier paid subscription.

− Recorded music revenues in millions \$ (adjusted for inflation, 2018 Dollars) of CD as victim technology (*V*)

*Remark*: values are at recommended or estimated list price (cf., RIAA, 2019).

These measures of technology can indicate the pathway of the evolution of technology in a context of competition in markets.

- *Model and data analysis procedure*
Model [5] of the killer technology is specified as follows:

$$Log\ Kl_t = log a + B\ log\ V_t + u_t \qquad [6]$$

*a* is a constant; *log* has base *e*= 2.7182818; *t*=time; $u_t$ = error term

$Kl_t$ is a measure of the growth of killer technology in markets

$V_t$ is a measure of the growth of victim technology in markets

The equations of simple regression [6] are estimated using the Ordinary Least Squares method. Statistical analyses are performed with the Statistics Software SPSS® version 24.

**Results**

☐ *Case study A: farm tractor technology*

**Table 1** – Parametric estimates of the model of killer technology based on farm tractor technology, 1920-1960 period in U.S. market

| *Dependent variable*: *log* number of tractors on farms in thousands as killer technology (mechanical power) | | | |
|---|---|---|---|
| | *Constant* $\alpha$ *(St. Err.)* | *coefficient* $\beta$=B *(St. Err.)* | $R^2$ *adj.* *(St. Err. of the Estimate)* | *F (sign.)* |
| Farm tractor | 20.36*** (0.69) | −1.42*** (0.08) | 0.90 (0.26) | 352.20 (0.001) |

*Note*: *** significant at 1‰; Explanatory variable is *log* Number of horses on farms in thousands as victim technology (animal power)

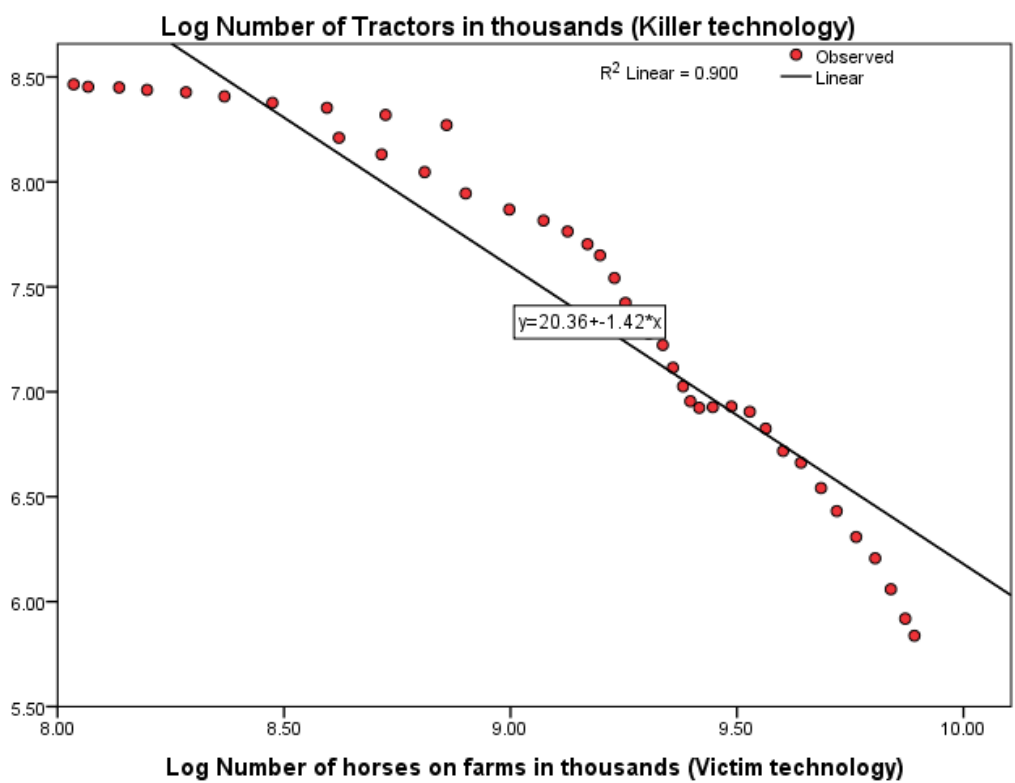

**Figure 1.** Trend and estimated relationship of the growth of farm tractor technology, 1920-1960 period in U.S. market

□ *Case study B: thermoelectric generating units*

**Table 2** – Parametric estimates of the model of killer technology based on thermoelectric generating units, 1917-1972 period in Canadian market

| *Dependent variable*: *log* total installed capacity of thermal power in MW as killer technology | | | |
|---|---|---|---|
| *Constant* <br> *α* <br><br> *(St. Err.)* | *Coefficient* <br> *β*=B <br><br> *(St. Err.)* | *R² adj.* <br> *(St. Err.* <br> *of the* <br> *Estimate)* | *F* <br> *(sign.)* |
| Thermoelectric generating units | −6.06*** | 1.46*** | 0.87 | 358.64 |
| | (0.69) | (0.08) | (0.53) | (0.001) |

*Note*: *** significant at 1‰; Explanatory variable is *log* total installed capacity of hydroelectric power in MW as victim technology

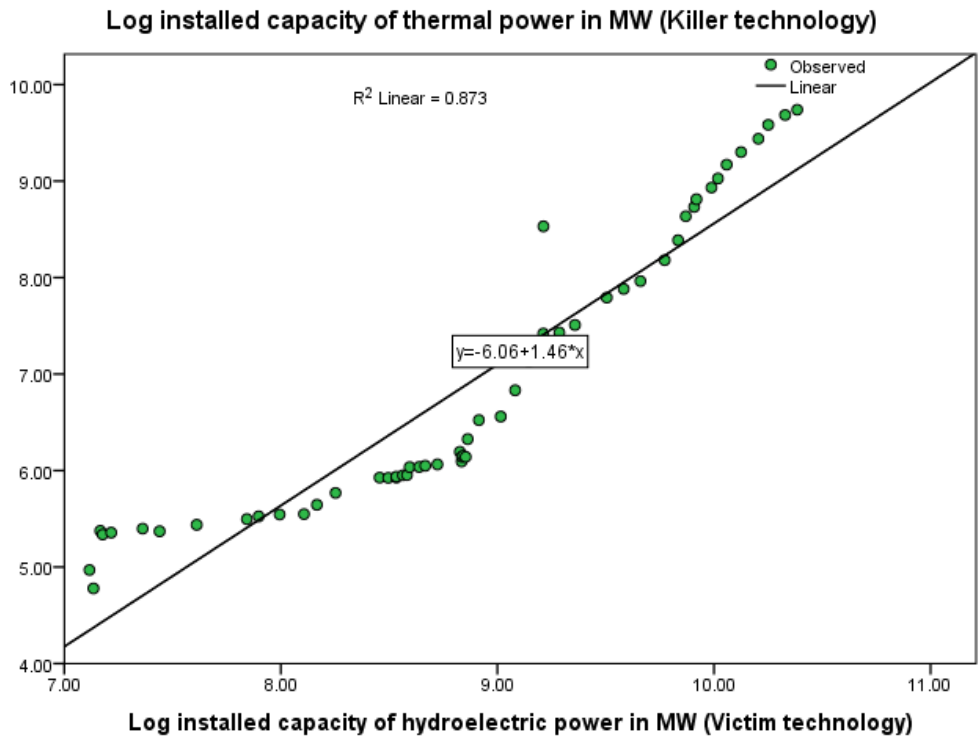

**Figure 2.** Trend and estimated relationship of the growth of thermoelectric generating units, 1917-1972 period in Canadian market

□ *Case study C: diesel-powered tractors*

**Table 3** – Parametric estimates of the model of killer technology based on diesel-powered tractors, 1955-1971 period in U.S. market

| *Dependent variable*: *log* annual production of diesel-powered tractors as killer technology | | | | |
| --- | --- | --- | --- | --- |
| | *Constant* $\alpha$ (*St. Err.*) | *Coefficient* $\beta$=B (*St. Err.*) | $R^2$ adj. (*St. Err. of the Estimate*) | *F (sign.)* |
| Diesel-powered tractors | 20.21*** | −0.76*** | 0.52 | 18.24 |
| | (2.08) | (0.18) | (0.38) | (0.001) |

*Note*: ***significant at 1‰; Explanatory variable is *log* production of gasoline-powered tractors as victim technology

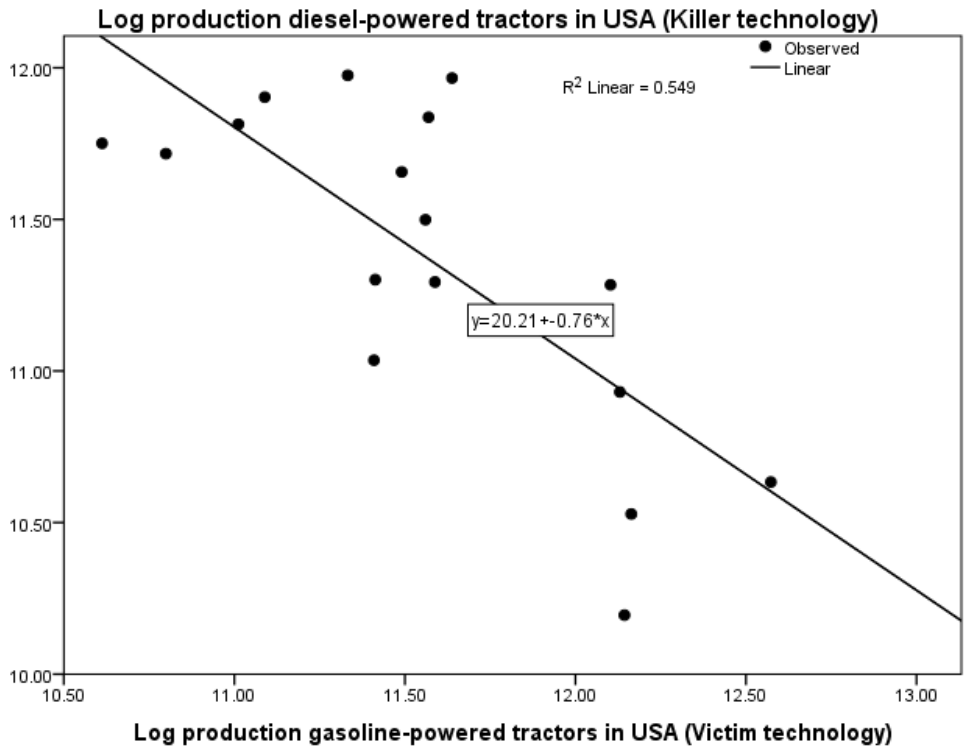

**Figure 3.** Trend and estimated relationship of the growth of production of diesel-powered tractors, 1955-1971 period in U.S. market

The parametric estimated relationships in Tabb. 1-3 and represented in Figs. 1-3 show that the significance of the coefficients and the explanatory power of equations are very high. The $R^2$ adj. is also very high and two models explain more than 85% variance in the data, whereas model of diesel-powered tractors explains more than 50% of variance.

The results show that the relative growth rate (measured with coefficient of regression) of killer technology is significantly different from unity, indicating that the disruption of new technology for the other one generally involves a process of disproportionate growth of one in relation to the other. In particular, results suggest that farm tractor in the USA with $B=$ $-1.42$ (i.e., <1) destroys at a lower relative rate of change the animal power of horses in agricultural operations. The diesel-powered tractors in the United States have also, over 1955-1971, a negative coefficient $B=-0.76$, such that this technology substitutes (i.e., kills) gasoline-powered tractors at lower rate of change. Finally, the competition between hydro and thermoelectric power in Canada over 1917-1972 has $B=1.46$ that is >1, suggesting that thermoelectric generating units destroy hydroelectric power at a greater relative rate of change over the course of time (*development of killer technology*).

□ *Case study D: Technology in recorded music with competition of cassette vs. CD and of CD vs. streaming technology*

An interesting case study is recorded music industry in the United States. From 1973 to 2018, the technological trajectories for delivering sound−included music−have had radical changes. In particular, during the 1970s and 1980s, the most common technological device to deliver music was compact cassettes based on analog magnetic tape for audio recording and playback. This product innovation was developed by Philips company, released in 1962 and introduced in the USA in 1964. Engineers began to work on techniques to increase the sound quality of cassette tapes, such as Ray Dolby that developed in 1968 a technology called Dolby noise reduction. These technological advances associated with cheaper prices and a higher performance of cassette than 8-track tapes (a tape cartridge introduced by William Lear in 1965 to be used in cars) led cassette tapes to be a dominant technology on 8-track tapes in the mid-1970s and in the early 1980s.

However, the emerging technology of compact audio disc (CD) co-developed by Philips and Sony and launched in 1982 generates a market shift (BBC News, 2007). CD is a digital optical disc data storage format originally developed to store and play only sound recordings but it was later adapted also for storage of other data (Coccia, 2018b). In the mid-1990s and in the early 2000s the sound quality of CD led this technology to be the dominant one in market, overtaking cassette sales from 1991 to 2005 (RIAA, 2019).

The revolution of Information and Communication Technologies (ICTs) has generated other new technologies for market of recorded music, based on transmission of video/audio information over the Internet, such as:

*Download mode.* The content file is completely downloaded and then played. This mode requires long downloading time for the whole content file and needs a hard disk space.

*Streaming mode.* The content file is not required to be downloaded completely and it is playing while parts of the content are being received and decoded.

In particular, the video streaming technology delivers audio and video over the Internet to reach many customers using their personal computers, personal digital assistants, mobile smartphones or other streaming devices. The growth of streaming technology is due to broadband networks, efficient techniques of video and audio compression, a higher quality and variety of audio and video services over Internet. A streaming media player can be either an integral part of a browser, a plug-in, a separate program, or a dedicated device, such as Apple TV, iPod, etc. For streaming technology UDP/IP (User Datagram Protocol/ Internet Protocol) is used to deliver the multi-media flow as a sequence of small packets. The application of layer protocol RTP/RTSP (Real-time Transport Protocol /Real Time Streaming Protocol), which is implemented on top of UDP/IP, provides an end-to-end network transport for video streaming.

There are many modes of streaming video content distribution (cf., RIAA, 2019):

- *Sound Exchange Distributions* based on payments to performers and copyright holders for digital radio services under statutory licenses
- *Paid Subscription* includes streaming, tethered, and other paid subscription services not operating under statutory licenses
- *Limited Tier Paid Subscription* includes streaming services with interactivity limitations by availability, device restriction, catalog limitations, on demand access, or other factors
- *On-Demand Streaming* includes Ad-supported audio and music video services not operating under statutory licenses
- *Other Ad-supported Streaming* includes revenues paid directly for statutory services that are not distributed by Sound Exchange and not included in other streaming categories.

This case study focuses on a period in which there are data of technologies in competition.

*Phase 1. CD as killer technology of Cassette technology*

**Table 4**  Parametric estimates of the model of killer technology based on CD technology, 1984-2008 period in U.S. market

| | | | | |
|---|---|---|---|---|
| *Dependent variable: log* annual recorded music revenues of CD (value adjusted for inflation, 2018 dollars) as killer technology | | | | |
| | *Constant* | *Coefficient* | *R² adj.* | |
| | *α* | *β=B* | *(St. Err.* | *F* |
| | | | *of the* | *(sign.)* |
| | *(St. Err.)* | *(St. Err.)* | *Estimate)* | |
| CD technology | −9.8* | 2.1*** | 0.51 | 14.38 |
| | (4.72) | (0.55) | (0.64) | (0.003) |

*Note*: *** significant at 1‰; * significant at 1%; Explanatory variable is *log* annual recorded music revenues of cassette (value adjusted for inflation, 2018 dollars) as victim technology

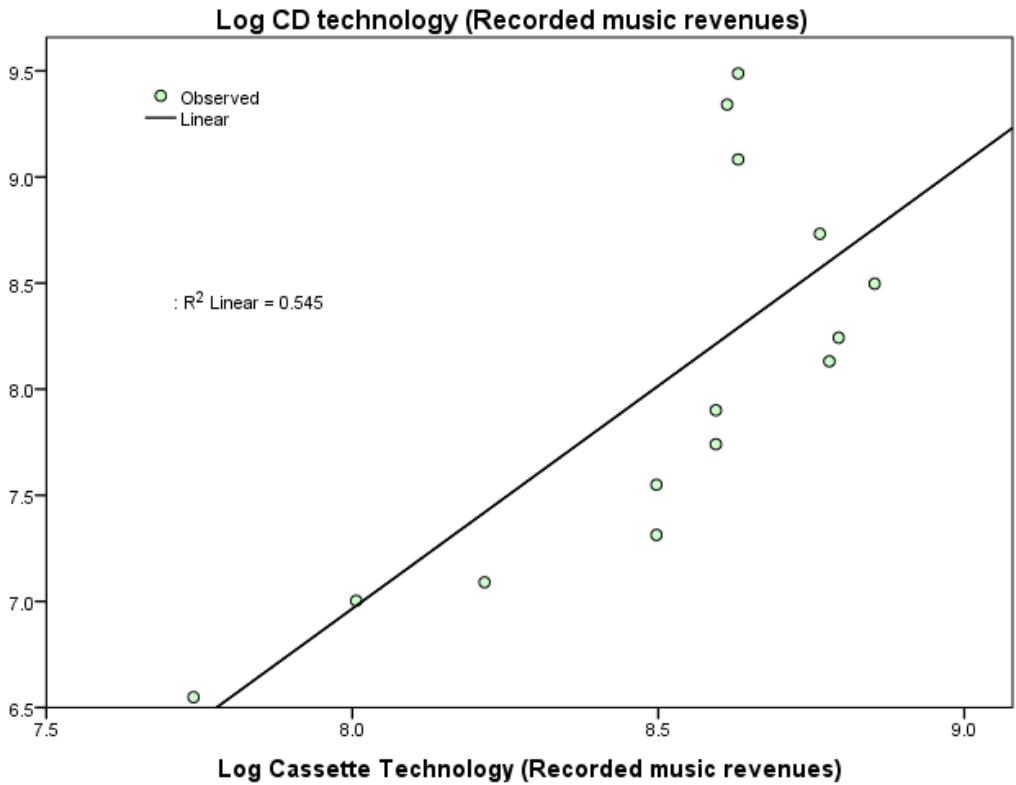

**Figure 4.** Fit line and estimated relationship of the growth of recorded music revenues (value adjusted for inflation, 2018 dollars) of CD technology (*killer technology*) on Cassette technology (*victim technology*), 1984-2008 period in U.S. market (log scale)

*Phase 2. Streaming technology as killer technology of CD technology (2004-2018)*

**Table 5** Parametric estimates of the model of killer technology based on streaming technology, 2004-2018 period in U.S. market

| | Constant $\alpha$ (St. Err.) | Coefficient $\beta=B$ (St. Err.) | $R^2$ adj. (St. Err. of the Estimate) | F (sign.) |
|---|---|---|---|---|
| *Dependent variable: log* annual recorded music revenues of streaming technology (value adjusted for inflation, 2018 dollars) as killer technology | | | | |
| Streaming technology | 17.22*** | −1.28*** | 0.95 | 240.01 |
| | (0.67) | (0.08) | (0.27) | (0.001) |

*Note*: *** significant at 1‰; Explanatory variable is *log* annual recorded music revenues of CD technology (value adjusted for inflation, 2018 dollars) as victim technology.

Note that streaming technology is measured here including recorded music revenues of different modes: paid subscription, on-demand streaming, other Ad-supported streaming, sound exchange distributions and limited tier paid subscription.

The parametric estimated relationship in Tab. 4 and represented in Fig. 4 shows that the significance of the coefficients and the explanatory power of equation are high. The $R^2$ adj. is also high and model of CD technology as killer technology on Cassette technology explains more than 50% variance in the data (Tab. 4). The results show that that CD technology in the USA with *B*= 2.1 (i.e., >1) has destroyed at a high relative rate of change the market of cassette technology (period 1984-2008).

Tab. 5 and Fig. 5 show results of the second phase under study based on a shorter period (from 2004 to 2018 =14 years). Streaming technology in this period is still in the phase of development, such that it is destroying CD technology in markets at a lower rate of change (*B*=−1.28, that is <1).

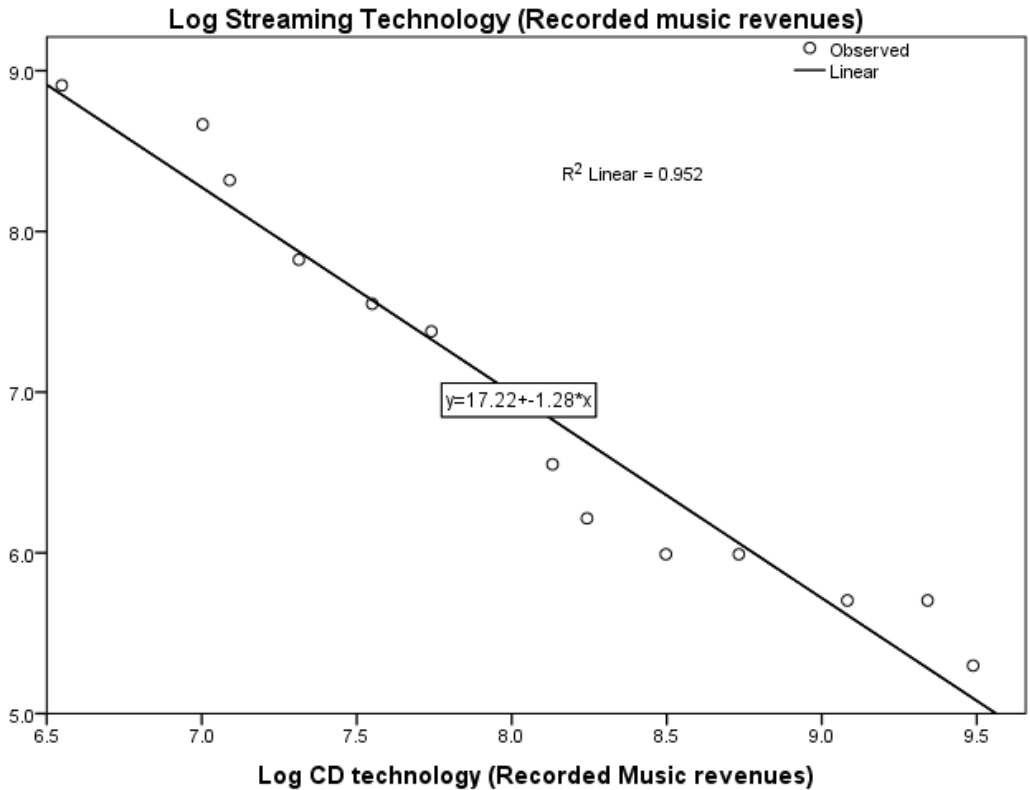

**Figure 5.** Fit line and estimated relationship of the growth of recorded music revenues (value adjusted for inflation, 2018 dollars) of Streaming technology on Cassette technology, 2004-2018 period in U.S. market (log scale).

Note that streaming technology is measured here including different modes: paid subscription, on-demand streaming, other Ad-supported streaming, sound exchange distributions and limited tier paid subscription.

The study shows that the U. S. recorded music revenues of streaming technologies have overtaken CD technology in 2015 with $2,400 millions vs. $1,400 millions. This short run of data analyzed (i.e., 2015-2018) justifies a lower rate of change with which this killer technology (i.e., streaming technology in the initial phase of development) is destroying the victim technology of CD. Instead, in the first phase, the long term period of substitution of CD technology on Cassette technology, started in 1991 with recorded music revenues of $4,300 million of CD vs. $3,000 millions of Cassette (about 17 years recorded, from 1991 to 2008), it explains the high rate of substitution of CD as killer technology on Cassette as victim technology.

o   *Theoretical and empirical laws of killer technology in the market of recorded music*

The analysis of recorded music industry shows the evolution of different product innovations and new technologies given by (cf. Tab. 6):

*8-track* (1965-1982)⟹*cassette* (1964-2005)⟹*CD* (1983-2018)⟹*Download technology* (2004-in progress) ⟹*streaming technology* (2005-in progress)

**Table 6** – Average period of killer technology to destroy more than 50% of total revenue of established technology in recorded music industry of U.S. market

| Established Technology in market of recorded music | Year of the introduction of <u>established</u> technology | *New* killer technology in market of recorded music | Year in which new technology destroys more than 50% of the revenue of <u>established</u> technology | % of recorded music revenues of <u>established</u> technology | Peak of revenues (<u>established</u> technology) | Ending of revenues (established <u>technology</u>) | Disruption Period (*DP* in years) of <u>established</u> technology *via* new killer technology |
|---|---|---|---|---|---|---|---|
| | | | | | M | Z | ***DP*=Z−M** |
| 8-track | 1965 | *Cassette* | 1980 | 42.80 in 1980 | 1978 | 1982 | **4** |
| Cassette | 1964 | *CD* | 1991 | 41.00 in 1991 | 1990 | 2005 | **15** |
| CD | 1983 | *Download* | 2012 | 45.20 in 2012 | 2001 | 2018 | **17** |
| CD | 1983 | *Download+ Streaming* | 2011 | 46.60 in 2011 | | | |
| Download | 2004 | *Streaming* | 2015 | 49.98 in 2015 | | | |
| | | | *Average values* | 45.12% | *Average values* | | 12 years |
| | | | *Standard Deviation (SD)* | 3.47% | *Standard Deviation (SD)* | | 7 years |

*Note*: elaboration on data by RIAA (2019);
Disruption period of established technology is MZ = year with ending of revenues − year with the peak of revenues.

The analysis of this market with the perspective of killer technologies suggests the following theoretical and empirical laws within technological change:

o The first economic law of technological disruption states that a new technology destroys the established technology, overtaking the percentage of total revenue in market, in an average period of 12 years (SD=7years).

   *Proof.*

   Table 6 is the empirical evidence of this law, showing the average duration of disruption phase of killer technologies on victim technologies in recorded music market.

The analysis of data by RIAA (2019) also shows different technological waves driven by different radical innovations introduced in U.S. recorded music market.

The *first technological wave* is due to 8-track tape introduced by William Lear in 1965 (using previous technology of tape cartridge introduced in 1958 by the Radio Corporation of America-RCA-Records Label) to be used in cars and supported by a growing automotive industry. The peak of 8-track tape measured with U.S. recorded music revenues is achieved in 1978 (RIAA, 2019), after 13 years of its introduction. However, in 1964 is also introduced in U.S. recorded market the cassette technology developed by Phillips company. This new technology has destroyed 8-track tape in 1982 with a disruption period of 4 years, given by difference between year with ending of revenues of 8-track tape and year with the peak of revenues (i.e., 1982 −1978=4 years; cf., Tab. 6). The length of technological cycle of 8-track tape is 17 years (from 1965 to 1982; cf., Tab. 7).

The *second technological wave* is due to cassette technology that started in U.S. recorded music market in 1964 and achieved the peak in 1990. However, it is destroyed by killer technology of CD in 2005. The length of technological cycle of cassette technology is about 41 years, given by difference between year with the starting of revenues − year with the ending of revenues (cf., Tab. 7).

The *third technological wave* is by CD that achieves the peak in 2001, after 18 years from its introduction in 1983. In 2018 this technology is almost destroyed by new technologies of download and video streaming. CD technology in 2018 has a mere $698.4 million of revenue on a total of $9,846 million in U.S. recorded music market. The length of technological cycle of CD technology is about 35 years, whereas the disruption period is about 17 years (cf., Tabb. 6-7).

The *on-going fourth technological wave* of recorded music market is due to download and streaming technology introduced in the mid-2000s. However, download mode has had the peak in 2012, after 8 years from its introduction in 2004, and now it has a phase of decline because of streaming technology that is growing, driven by many technical advances, growing video-sharing websites and general advantages for consumer use (cf., Tab. 7).

**Table 7** – Technological cycles in the U.S. recorded music industry

| | Technological wave in U.S. recorded music market | Upwave | | Downwave | AM length upwave years =M-A | MZ length downwave years [1] =Z-M | AZ length cycle years =Z-A | (M / AZ % | MZ / AZ % |
|---|---|---|---|---|---|---|---|---|---|
| | | A begin of revenues | M peak of revenues | Z end of revenues | | | | | |
| 1 | 8-track tape technology | 1965 | 1978 | 1982 | 13 | 4 | 17 | 76.47 | 23.53 |
| 2 | Cassette technology | 1964 | 1990 | 2005 | 26 | 15 | 41 | 63.41 | 36.59 |
| 3 | CD technology | 1983 | 2001 | 2018 | 18 | 17 | 35 | 51.43 | 48.57 |
| 4 | Download technology | 2004 | 2012 | * | 8 | - | - | - | - |
| 4 | Streaming technology | 2005 | * | * | - | - | - | - | - |
| | *Arithmetic mean years* | | | | 19.00 | 12.00 | 31.00 | 63.77% | 36.23% |
| | *Standard Deviation (SD) years* | | | | 6.56 | 7 | 12.49 | | |

*Note.* * is a technology in progress; elaboration on data from RIAA (2019);

(1) Disruption period of established technology is MZ = year with ending of revenues of technology − year with the peak of revenues; length of technological cycle of technology is AZ= year with the starting of revenues − year with the ending of revenues.

These empirical results in Tab. 7 suggest other empirical law for killer technologies:

o The second law states that upwave of technological cycle is longer than downwave phase (asymmetric path of technological cycle).

*Proof.*

The analysis of three technological waves (8-track tape, cassette and CD technology) shows that upwave has an average duration of 19 years (SD=6.56y), whereas downwave phase has an average duration of 12 years (SD=7y). Average duration of technological cycle in recorded music market is about 31 years (Tab. 7). In particular, results show that technological cycles have an average upwave duration equal to 63.77% of wavelength, whereas the average downwave duration is shorter: about 36.23% of overall wavelength.

*Remark.* Coccia (2010) showed that economic long waves have not a symmetric and regular dynamics but they have asymmetric paths with longer periods of *upwave* than *downwave* over time.

o The third law states that killer technology destroys faster established technology when the period of introduction of killer technology is close to the introduction period of established technology.

*Proof.*

8-track tape is introduced in 1965, whereas cassette technology in 1964 (1 year before) such that killer technology of cassette has destroyed 8-track technology in about 4 years from its peak of revenue (see Tabb. 6-7). The comparison of CD *versus* cassette shows that CD is introduced in about 1983, after 19 years from cassette (introduced in 1964). CD technology has destroyed cassette in 15 years. *Mutatis mutandis,* download and streaming technologies are both introduced in recorded music market in 2004-2005; results show that streaming technology as killer one is destroying very fast the download mode in U.S. recorded music market. In 2018, download mode has $1,037 million revenue accounted for 10.53% of total revenue ($9,846), whereas streaming technology has about 75% of total revenues in U.S recorded music market (i.e., $7,367 on a total of $9,846).

The findings here can be explained with the critical role of killer technologies that have technical and economic performance higher than other established technologies, generating a destruction of other technologies in markets. This technological behavior can be due to ambidexterity learning processes of killer technology, given by:

– "*learning via diffusion*" (Sahal, 1981, p. 114, Italics added) in which the increased adoption of a technology supports the path for improvement in its technical characteristics (i.e., technological advances).

– "*diffusion by learning*" that improvement in the technical characteristics of a technology enhances the scope for its adoption over the course of time (Sahal, 1981, p. 114, Italics added).

**DISCUSSION**

The concept of competition is frequently used to explain the diffusion and evolution of innovation and technology in industrial economics (Fisher and Pry, 1971; Porter, 1980; Utterback et al., 2019). The competition between technologies leads to a process of disruptive creation that generates technological and economic change over time (Calvano, 2006). In particular, a vital radical innovation in the dynamics of disruptive creation is the killer technology that may explain and generalize characteristics of the competition between technologies that generates competitive advantage of firms and technical change in society. Killer technology tends to affect the behavior of other technologies, generating in the long run a process of actual substitution of a new technique for the old (victim technology), and as a consequence, technical change in socioeconomic systems.

The third goal of this study stated in the introduction is to suggest the properties of killer technologies in industrial competition, based on proposed theoretical framework and empirical evidence, given by:

1. The nature and significance of killer technology is always associated with some comparable established technology in markets

2. The growth of killer technology is generally an allometric process of growth given by a disproportionate growth of killer technology in relation to the victim technology.

3. In the short run, killer technology can induce incremental technological advances of established technologies that have a prospect of being supplanted by a (new) killer technology.

4. In the long run, killer technology has a series of technological advances of its own resulting from various major and minor innovations to pave the way for the dominance over other established technologies in markets.

5. The long-run behavior and evolution of any killer technology is not independent of the behavior of other inter-related technologies.

6. The ambidexterity learning processes based on *learning via diffusion* and *diffusion by learning* are a driver underlying the development and adoption of killer technology versus victim technology in turbulent (complex and fast changing) markets.

7. The competition between killer technology and victim technology is a function of their inter-related patterns of growth and environment with socioeconomic, political and institutional change (Coccia, 2019a)[2].

The study documented here makes a unique contribution, for the first time to our knowledge, by showing the behavior and characteristics of a critical radical innovation (*killer technology*) in the process of creative disruption and how these killer technologies compete with established technologies to achieve the dominance in markets and generate technical change in society.

The theory here suggests a simple model that can predict the degree and rate at which *killer technologies* are adopted when they attempt in substituting for existing *victim technologies*.

These results suggest general properties that can support innovation strategy of firms on critical decisions of when to invest in R&D of new killer technologies, abandon the old technology or pursue an intermediate level of R&D investment between old and new technology for sustaining and safeguarding competitive advantage in turbulent markets.

In general, the study here suggests a theoretical framework to explain one of the characteristics of the competition between technologies that generates technological change in society.

However, the idea of a *killer technology* in markets is adequate in some cases but less in others because of the diversity of technologies in socioeconomic ecosystems (cf., Coccia, 2018, 2019b; Pistorius and Utterback, 1997). Nevertheless, this study keeps its validity in explaining and predicting several phenomena of the competition between technologies in turbulent markets with final dominance of vital technologies that generate technical, economic and social change (cf., Berg et al., 2019; Grodal et al., 2015; Kauffman and Macready, 1995, p. 27ff).

---

[2] Patterns of technological innovation are affected by manifold factors such as R&D investment, level of democracy, predominant religion, growth rate of population, etc. (cf. studies by Calabrese et al., 2005; Cavallo et al., 2014; Coccia, 2005, 2009, 2010a, 2011, 2012, 2013, 2014, 2014a, 2014b, 2014c, 2015, 2015a, 2015b, 2016, 2016a, 2017a, 2018c, 2018d).

The theoretical framework of killer technology is a reasonable starting point for understanding universal drivers of technical change, though theory here—of course—cannot predict any given behavior and characteristics of specific technologies in technological change. We know, *de facto*, that other things are often not equal over time and space in the domain of technology.

Overall, then, the proposed theory here may lay the foundation for development of more sophisticated concepts and theoretical frameworks in economics of innovation. These findings here, *de facto*, can encourage further theoretical exploration in the *terra incognita* of the competition between technologies that generates disruptive creation for technological and economic change in society. Future efforts in this research field will be directed to provide further empirical evidence, also considering dependency-network framework between technologies to better explain the nature and behavior of killer technologies in markets (cf., Mazzolini et al., 2018; Iacopini et al., 2018). To conclude, identifying a generalizable theory to explain the behavior and characteristics of new typologies of technological innovation within industrial competition is a non-trivial exercise. In fact, Wright (1997, p. 1562) properly claims that: "In the world of technological change, bounded rationality is the rule."